\documentclass[12pt]{article}
\usepackage[top=30truemm,bottom=30truemm,left=27truemm,right=27truemm]{geometry}

\usepackage{indentfirst}
\usepackage[dvipdfmx]{graphicx}
\usepackage{float}
\usepackage{url}
\usepackage{siunitx}
\usepackage{array}
\usepackage{amsmath}
\usepackage{amssymb}
\usepackage{fancybox}
\usepackage{ascmac}
\usepackage{bm}
\usepackage{color}
\usepackage{authblk}
\usepackage{appendix}
\usepackage{caption,latexsym,amsfonts,mathtools,here}
\usepackage{hyperref}
\usepackage[sorting=none,doi=false]{biblatex}

\bibliography{main}

\numberwithin{equation}{section}

\newcommand{\no}{\nonumber}
\newcommand{\h}{\hspace{1mm}}
\newcommand{\hhh}{\hspace{6mm}}
\newcommand{\ds}{\displaystyle}

\newcommand{\pa}{\partial}

\newcommand{\vare}{\varepsilon}
\newcommand{\RRa}{R^{0}{}_{i0m}R^{0m}{}_{jk}}
\newcommand{\RRb}{R^{0}{}_{ilm}R^{lm}{}_{jk}}

\begin{document}

\begin{titlepage}
\begin{flushright}
KOBE-COSMO-23-04  
\end{flushright}

\vspace{28pt}

\begin{center}

{\large{\textbf{Circularly polarized gravitational waves in Chern-Simons gravity\\
originated from an axion domain wall}}}

\vspace{50pt}

{Sugumi Kanno$^*$, Jiro Soda$^{\flat, \sharp}$, and Akira Taniguchi$^*$}
\end{center}

\vspace{20pt}

\shortstack[l]
{\hspace{1cm}\it $^*${\small Department of Physics, Kyushu University, Fukuoka 819-0395, Japan}\\[4pt]
\it \hspace{1cm}$^\flat${\small Department of Physics, Kobe University, Kobe 657-8501, Japan}\\[4pt]
\it \hspace{1cm}$^\sharp${\small International Center for Quantum-field Measurement Systems for Studies}\\[4pt]
\it {\hspace{1.15cm}of the Universe and  Particles (QUP), KEK, Tsukuba 305-0801, Japan}}

\vspace{28pt}

%==========================================================
\begin{abstract}
      We study a scattering problem of gravitational waves (GWs) by an axion domain wall in Chern-Simons (CS) gravity. We find that circular polarization of GWs is produced after passing through the domain wall. It turns out that the circular polarization is sizable if the frequency of the GW is comparable to a critical value determined by the characteristic CS length scale and the energy scale of the axion domain wall. Thus, observations of the circular polarization  could  give a stringent constraint on the characteristic CS length scale or could be a new avenue to search for axion.
\end{abstract}
%==========================================================
\end{titlepage}

\tableofcontents

\section{Introduction}

Gravitational waves (GWs) are a useful probe of new physics. 
In string theory, axion ~\cite{Kim:1979if,Shifman:1979if,Zhitnitsky:1980tq,Dine:1981rt} or axion-like particles (ALPs)~\cite{Arvanitaki:2009fg} are ubiquitous. We call the ALPs axion hereafter. Since the axion is pseudo-scalar, it typically couples with GWs through Chern-Simons (CS) term~\cite{Campbell:1990fu,Jackiw:2003pm}. If the axion has an expectation value, the CS term often induces circular polarization of GWs~\cite{Lue:1998mq}. 
For example, axion inflation in CS gravity can produce circularly polarized primordial GWs~\cite{Satoh:2008ck}, and the circular polarization is shown to be enhanced in the presence of the Gauss-Bonnet term~\cite{Satoh:2007gn}. 
Axion is also a candidate for dark matter~\cite{Preskill:1982cy,Abbott:1982af,Dine:1982ah}.
 Studies of circular polarization of GWs propagating in homogeneous axion dark matter are initiated in 
\cite{Yoshida:2017cjl,Chu:2020iil} but it was pointed out that the circular polarization is not large enough by using LIGO data or in a realistic halo profile by subsequent studies ~\cite{Jung:2020aem,Fujita:2020iyx,Tsutsui:2023jbk}. In this paper, we investigate circular polarization of GWs passing through a domain wall which consists of the axion. In this case, the axion becomes inhomogeneous in the background in contrast to the cosmological homogeneous axion studied so far. 
We analyze a scattering problem and study how
the axion domain wall induces the circular polarization of GWs.

 The formation of domain walls in the early universe is an important phenomenon~\cite{vilenkin1994cosmic}. When the Peccei-Quinn (PQ) symmetry is spontaneously broken after inflation, domain walls are formed~\cite{Sikivie:1982qv}. The energy density of the domain walls formed in such scenarios is of the order of $10^{9}\sim 10^{12}~\si{GeV}$\cite{Hiramatsu:2012gg,Graham:2015ouw}. However, since their energy density decays too slowly relative to the energy density of the surrounding matter, they overclose the universe~\cite{Zeldovich:1974uw}. 
 This is a notorious domain wall problem.
 On top of the domain wall problem, CMB observations give a stringent constraint on the energy density of domain walls $\sigma < (0.93\ \si{MeV})^3 $ at the 95\% confidence level for the standard $\Lambda$-CDM cosmology~\cite{Lazanu:2015fua}.
 
 There are several ideas to resolve the domain wall problem. One can simply consider pre-inflationary breakdown of the PQ symmetry. In~\cite{Larsson:1996sp}, the authors consider a tilt of the potential or biased initial conditions.
Interestingly, in the scenario \cite{Babichev:2021uvl}, the domain wall does not have the domain wall problem because the energy density of the domain wall decreases faster than that of radiation. 

From the theoretical point of view, the coupling constant between axions and CS gravity is a free parameter. On the other hand, from the observational point of view, the strength of the coupling can be constrained by measurements of the frame-dragging of objects orbiting the Earth\cite{Smith:2007jm} or by observations of double-binary pulsars\cite{Yunes:2008ua}. Currently, the upper bound of the characteristic CS length scale $\ell$ is obtained as $\ell \lesssim 1 {\rm AU}\sim10^8$ km, which comes from measurements of frame-dragging effects around the Earth by Gravity Probe B~\cite{Ali-Haimoud:2011zme}.

  In this paper, we assume the existence of the axion domain wall within the cosmological horizon at present and evaluate the circular polarization of GWs when they pass through the domain wall. Then, we argue the observability of the circular polarization in this setup, and discuss the implications of circular polarization for the coupling constant between the axion and CS gravity. If the circular polarization is observed, it would be a new avenue to search for axion.

The paper is organized as follows. In section 2, we present the setup and derive the equation of motion (EOM) of GWs in CS gravity. In section 3, we solve the EOM of GWs passing through the domain wall. In section 4, we discuss phenomenological implications. In particular, we argue how to constrain the coupling constant.  The final section is devoted to  conclusion. In the appendix, the derivation of basic equations is shown.
We work in the natural unit: $\hbar=c=1$.
%
%
%==========================================================

\section{GWs passing through a domain wall in CS gravity }

%==========================================================
In this section, we consider GWs propagating in the background of axion domain wall. We first present a background solution of the axion domain wall and then consider the GWs in the background.
\subsection{Axion domain wall background}
The action of the axion field $\phi$ is given by
\begin{eqnarray}
	S=-\int d^4x\sqrt{-g}\left[\frac{1}{2}\pa_\mu\phi\,\pa^\mu\phi+V(\phi)\right] \,,
 \label{eq:action}
	\end{eqnarray}
where $V(\phi)$ is a double-well potential with two minima located at $\phi = \pm\eta$ such as
%
	%potential of phi
	\begin{eqnarray}
	\label{eq:potential}
	V(\phi) = \frac{\lambda}{4} (\phi^2 - \eta^2)^2\,,
	\end{eqnarray}
Here, $\lambda$ is a coupling constant. Note that the action is parity even because of the quadratic form of the $\phi$ even if the $\phi$ is pseudo-scalar. We assume that the domain wall is static and planar. Then, without loss of generality, the planar domain wall is assumed to be in the $(x,y)$-plane and it is orthogonal to the $z$-axis. In this case, the corresponding domain wall solution is found to be
%
	%domain wall
	\begin{eqnarray}
	\label{domain wall}
	\phi (z) = \eta \tanh\left(\sqrt{\frac{\lambda}{2}}\eta z\right)\,.
	\end{eqnarray}
By using this solution, the surface energy density is calcualted as
%
	%domain wall
	\begin{eqnarray}
    \label{sigma}
	\sigma=\int dz\,\phi^{\prime\,2}(z) \sim \sqrt{\lambda}\,\eta^3\ .
	\end{eqnarray}
In the next subsection, we consider the situation
 in which GWs propagate in the $z$-direction and cross the domain wall. In our setup, we ignore the backreaction of the domain wall to the spacetime. 

\subsection{Gravitational waves in Chern-Simons gravity}
Let us consider GWs in the Minkowski space expressed by the tensor mode perturbation in the three dimensional metric:
%
	%metric
	\begin{eqnarray}
	\label{eq:metric}
	ds^2 = -dt^2 + (\delta_{ij} + h_{ij}) dx^i dx^j,
	\end{eqnarray}
where $\delta_{ij}$ are Kronecker delta. The metric perturbation $h_{ij}$ satisfies the transverse traceless gauge conditions $h^{i}{}_{j,i} = h^{i}{}_{i} = 0$. The indices $(i,j)$  run from 1 to 3, and $(1,2,3)=(x,y,z)$. 

The action we consider is Chern-Simons gravity expressed by
	\begin{eqnarray}
	\label{eq:action}
	S=\ds\frac{M^2_{\rm p}}{2}\int d^4x\sqrt{-g}R
 +\frac{M_{\rm p} \ell^2}{8}\int d^4x\sqrt{-g}\,\phi\, R\tilde{R},
	\end{eqnarray}
where $M^2_{\rm p} = 1/(8 \pi G)$, $\ell$ is a length scale characterizing the coupling strength between the axion field and gravity and $R\tilde{R}={1/2}\,\varepsilon^{\rho\sigma\alpha\beta}R^{\mu\nu}{}_{\alpha\beta}R_{\nu\mu\rho\sigma}$. Here, the four dimensional Levi-Civita tensor $\varepsilon^{\rho\sigma\alpha\beta}$ is defined by $\vare^{0123}=-1/\sqrt{-g}$. Note that the action is parity invariant because the parity-odd Chern-Simons gravity $R{\tilde R}$ couples with the pseudo-scalar $\phi$. 
Substituting the metric Eq.~(\ref{eq:metric}) into the action Eq.~(\ref{eq:action}), we obtain the action of quadratic form of $h_{ij}$,
%
	%quadratic action
	\begin{eqnarray}
	\label{eq:q_action}
	\overset{(2)}S
    =
    \ds\frac{M_{\rm p}^2 }{8}\int d^4x
    \left[
    \dot{h}^{ij}\dot{h}_{ij}-h^{ij,k}h_{ij,k}
    -\frac{\ell^2 }{M_{\rm p}}\varepsilon^{izk} \pa_z \phi 
   \left\{
            \ddot{h}^m{}_i\dot{h}_{km}
           +\dot{h}_{im,\ell}
           \left(h^{\ell}{}_{k}{}^{,m}
            -h^{m}{}_{k}{}^{,\ell}\right)
        \right\}
    \right] ,\quad
	\end{eqnarray}
where a dot denotes the derivative with respect to the time and $\varepsilon^{ijk}$ is a three dimensional Levi-Civita symbol. Note that $\phi$ depends only on $z$. The variation of the action with respect to $h_{ij}$ gives the equation of motion for GWs in CS gravity,
	\begin{eqnarray}
	\label{eq:eom_01}
	\Box h^{ij} = 
 \frac{\ell^2}{2M_{\rm p}}
 \varepsilon^{zik}\left[\phi'\Box+\phi''\pa_z\right]\dot{h}^j{}_{k}
   + \frac{\ell^2}{2M_{\rm p}}\varepsilon^{zjk}\left[\phi'\Box+\phi''\pa_z\right]\dot{h}^i{}_{k} \ ,
	\end{eqnarray}
where a prime denotes the derivative with respect to $z$. 
Since we assumed that GWs propagate along the $z$-axis, the wave vector is expressed as $k^{\mu}=(\omega,0,0,\omega)$ in the asymptotic region. We introduce the polarization tensors for the right-handed and left-handed circularly polarized modes as
%
	%polarization tensor
	\begin{eqnarray}
	e^{(R)}_{ij}=
	\begin{pmatrix} 
		  1 & i & 0 \\
		  i & -1 & 0  \\
		  0 & 0 & 0  \\
	\end{pmatrix},
	\hhh
	e^{(L)}_{ij}=
	\begin{pmatrix} 
		  1 & -i & 0  \\
		  -i & -1 & 0  \\
		  0 & 0 & 0 \\
	\end{pmatrix}.
	\end{eqnarray}
Note that we have relations
\begin{eqnarray}
    \varepsilon^{zpk} e^{(R)}_{ik} = i e^{(R)}_{ip}\ ,
    \qquad
    \varepsilon^{zpk} e^{(L)}_{ik} = -i e^{(L)}_{ip}\ .
\end{eqnarray}
Using these polarization tensors, the GWs can be decomposed into the amplitude and the polarization such as
%
	%h_{ij}
	\begin{eqnarray}
	\label{eq:hij}
	h_{ij}(t,z) = h_R(t,z) e^{(R)}_{ij} + h_L(t,z) e^{(L)}_{ij}\ ,
	\end{eqnarray}
where $h_R(t,z)$ and $h_L(t,z)$ are the amplitude of the right-handed and left-handed modes respectively. Substituting Eq.~(\ref{eq:hij}) into Eq.~(\ref{eq:eom_01}), we obtain
%
	%eom of h_{R/L}
	\begin{eqnarray}
	\label{eq:eom_02}
	\Box h_{R/L}(t,z) 
 = \pm i  \frac{\ell^2}{M_{\rm p}}\left[\phi'\Box + \phi''\pa_z\right]\dot{h}_{R/L}(t,z)\ .
	\end{eqnarray}
The $\pm$ correspond to the right-handed and the left-handed mode, respectively. 
By using the time translation symmetry in Eq.~(\ref{eq:eom_02}), we can write them by the Fourier mode $h_{R/L}(t,z)=H_{R/L}(z)\,e^{i\omega t}$. 
Then Eq.~(\ref{eq:eom_02}) is written as
%
	%eom of H
 	\begin{eqnarray}
	\label{eq:eom_03}
	\left(1 \pm \frac{\omega\ell^2}{M_{\rm p}}
     \phi'\right)H''_{R/L} {\pm \frac{\omega\ell^2}{M_{\rm p}}}\phi''H'_{R/L} + \omega^2\left(1 {\pm \frac{\omega\ell^2}{M_{\rm p}}}\phi'\right)H_{R/L} = 0\ .
	\end{eqnarray}
Apparently, right-handed and left-handed modes obey different equations. Thus the circular polarization is expected to be produced after passing through the domain wall. Curiously,  the above equations have been derived in the context of condensed matter physics~\cite{Martin-Ruiz:2017cjt}.
%
%==========================================================

\section{Circular polarization induced by a domain wall}
In this section, we derive the effective potential for the GWs. Then we give a formula of reflection coefficients, transmission coefficients, and circular polarization.
\subsection{Effective potential for GWs}
\begin{figure}[t]
        \centering
        \includegraphics[width=11cm]{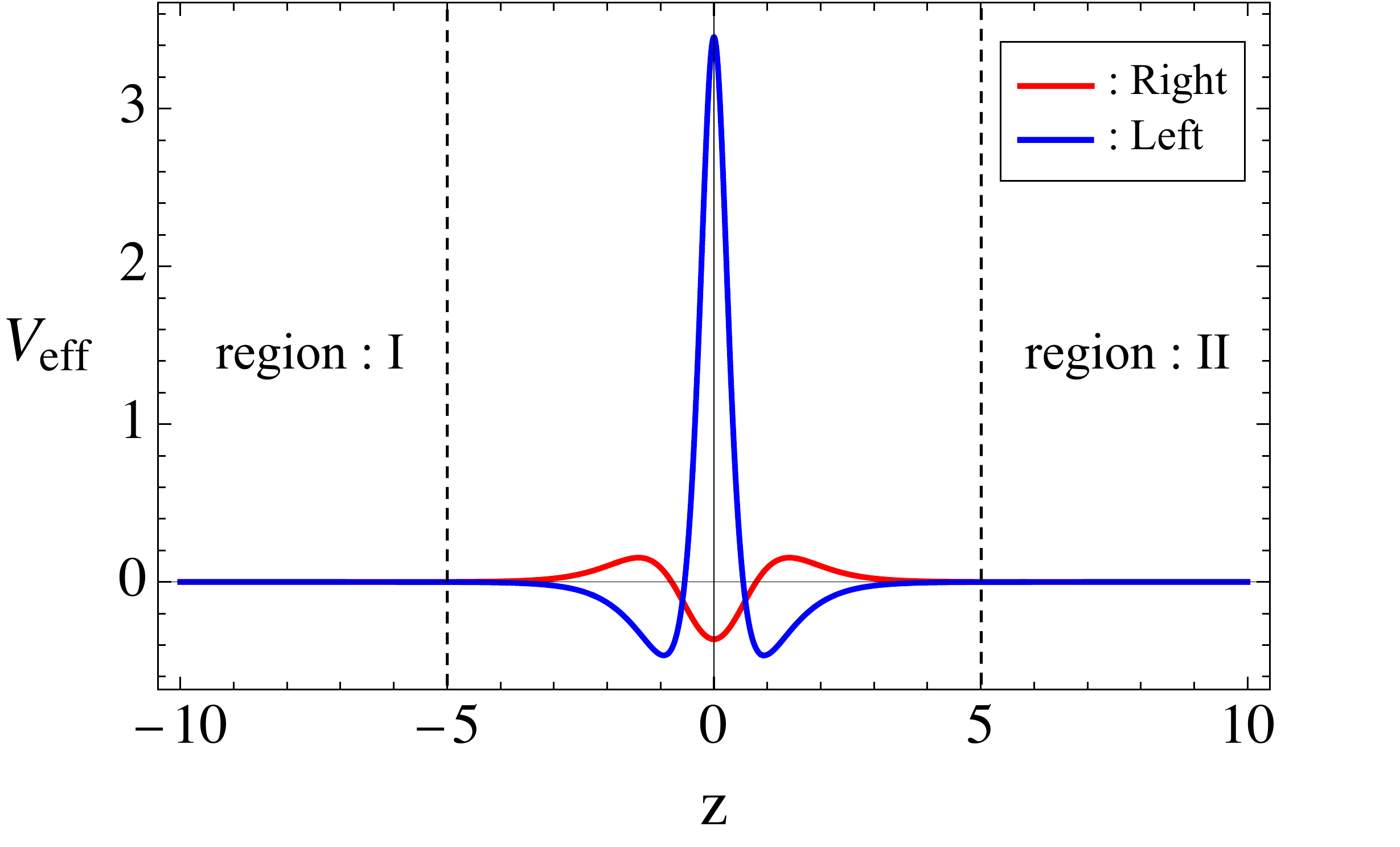}
        \vspace{-8pt}
        \caption{The effective potentials $V_{\rm{eff}}(z)$ are depicted for right-handed and left-handed circular polarization modes. The parameters are set to be $\eta = 0.9$, $\lambda =2$, and $\omega \ell^2/M_p$=1. We see 
        the potential vanishes in the asymptotic region I and II.}
        \label{fig:potential}
        \vspace{0.7cm}
    \end{figure}

If we change the variables in Eq.~(\ref{eq:eom_03}) such as
\begin{eqnarray}
F(z)\equiv 1\pm\frac{\omega\ell^2}{M_{\rm p}}\phi'(z)\,,\qquad
H_{R/L}(z)\equiv\frac{f_{R/L}(z)}{\sqrt{F(z)}}\,,
\end{eqnarray}
we obtain the canonical form of Eq.~(\ref{eq:eom_03}) given by
	%domain wall
	\begin{eqnarray}
    \label{eq:eom_sch}
	\left[-\frac{d^2}{d z^2} + V_{\rm{eff}}(z)\right]f_{R/L} = \omega^2 f_{R/L}\ ,
	\end{eqnarray}
where we can read off the effective potential $V_{\rm{eff}}(z)$ of the form
	\begin{eqnarray}
	\label{V_eff}
	V_{\rm{eff}} = -\frac{1}{4}\left(\frac{F'}{F}\right)^2+\frac{1}{2}\frac{F''}{F}   \ .
	\end{eqnarray}
The effective potential for right-handed and left-handed circular polarizations is depicted in Figure \ref{fig:potential} where the value of the potential vanishes to the far left (region:I) and far right (region:II). Thus, there is no distinction between $f_{R/L}(z)$ and $H_{R/L}(z)$ in these regions because $F(z)=1$.

\subsection{Reflection and transmission coefficients}
We now consider an incident wave from the region I which is scattered by the potential according to   
Eq.(\ref{eq:eom_sch}). Since the potential vanishes in the region I and II, we can express the solution of Eq.(\ref{eq:eom_sch})  as 
	\begin{eqnarray}
 \label{boundary}
	f_{R/L}(z)&=&
	\begin{cases}
   		 \ds\frac{e^{i\omega z}}{\sqrt{2\omega}}+\mathcal{R}_{R/L}\ds\frac{e^{-i\omega z}}{\sqrt{2\omega}} &\mbox{(region:I)}\\[8pt]
    		\mathcal{T}_{R/L}\ds\frac{e^{i\omega z}}{\sqrt{2\omega}} &\mbox{(region:I\hspace{-1pt}I)}\\
    	 \end{cases}\,,
	\end{eqnarray}
where $\mathcal{R}_{R/L}$ and $\mathcal{T}_{R/L}$ are the reflection and transmission coefficients. 
In order to obtain the transmission coefficient $\mathcal{T}_{R/L}$, we need to solve Eq.~(\ref{eq:eom_sch}) from the region $z<0$ numerically. However, the reflection coefficient ${\cal R}_{R/L}$ in $z<0$ is unknown a priori, so we consider two auxiliary problems. One is to consider an incident wave $e^{i\omega z}/\sqrt{2\omega}$ from the far left in the region I and a solution of Eq.~(\ref{eq:eom_sch}) in the region II. The other is to consider a wave  $e^{-i\omega z}/\sqrt{2\omega}$ that move backward in $z$ in the region I and a solution of Eq.~(\ref{eq:eom_sch}) in the region II. That is,
%
	%polarization
	\begin{eqnarray}
    \label{condition of H}
	f^{(+)}_{R/L}(z)&=&
        \begin{cases}
   		 \ds\frac{e^{i\omega z}}{\sqrt{2\omega}} &\mbox{(region\,:\,I)}\\[10pt]
    		A^{(+)}_{R/L}\ds\frac{e^{i\omega z}}{\sqrt{2\omega}} + B^{(+)}_{R/L}\ds\frac{e^{-i\omega z}}{\sqrt{2\omega}} &\mbox{(region\,:\,I\hspace{-1pt}I)}\\
        \end{cases}   \ ,   \\[10pt]
    f^{(-)}_{R/L}(z)&=&
        \begin{cases}
   		 \ds\frac{e^{-i\omega z}}{\sqrt{2\omega}} &\mbox{(region\,:\,I)}\\[10pt]
    		A^{(-)}_{R/L}\ds\frac{e^{i\omega z}}{\sqrt{2\omega}} + B^{(-)}_{R/L}\ds\frac{e^{-i\omega z}}{\sqrt{2\omega}}&\mbox{(region\,:\,I\hspace{-1pt}I)}   \\
        \end{cases}\ .
	\end{eqnarray}
We can construct the solution $f_{R/L}$ of the original problem Eq.~(\ref{boundary}) as a superposition of the solutions $f^{(+)}_{R/L}$ and $f^{(-)}_{R/L}$. If we impose the condition that no wave from the far right by multiplying it by $-B^{(+)}_{R/L}/B^{(-)}_{R/L}$, we can consider an incoming wave $e^{i\omega z}/\sqrt{2\omega}$ and a wave that moves backwards toward the far left $-\left(B^{(+)}_{R/L}/B^{(-)}_{R/L}\right)e^{-i\omega z}/\sqrt{2\omega}$. Then the reflection and transmisson coefficients in Eq.~(\ref{boundary}) can be obtained as
%
	%polarization
	\begin{eqnarray}
    \label{RT}
	\mathcal{R}_{R/L} = -\frac{B^{(+)}_{R/L}}{B^{(-)}_{R/L}}\ ,\hspace{0.6cm}\mathcal{T}_{R/L} =A^{(+)}_{R/L}-\frac{B^{(+)}_{R/L}}{B^{(-)}_{R/L}}A^{(-)}_{R/L}\ .
	\end{eqnarray}
Defining the mode function as $v(z)=e^{-i\omega z}/\sqrt{2\omega}$, we can deduce $A^{(\pm)}_{R/L}$ and $B^{(\pm)}_{R/L}$ as follows
%
	%polarization
	\begin{eqnarray}
	A^{(\pm)}_{R/L} & = & -i\left[v(z)f'^{\,(\pm)}_{R/L}(z)-f^{(\pm)}_{R/L}(z)v'(z)\right]\ ,\\[8pt]
    B^{(\pm)}_{R/L} & = & i\left[v^{*}(z)f'^{\,(\pm)}_{R/L}(z)-f^{(\pm)}_{R/L}(z)v^{*'}(z)\right]\label{B}\ ,
	\end{eqnarray}
where an asterisk denotes the complex conjugate. The degree of the circular polarization is defined as 
%
	%polarization
	\begin{eqnarray}
	\Pi \equiv \frac{|\mathcal{T}_R|^2-|\mathcal{T}_L|^2}{|\mathcal{T}_R|^2+|\mathcal{T}_L|^2}\ .
	\end{eqnarray}

%==========================================================

%==========================================================

\section{Implications for cosmology}

\begin{figure}[t]
        \centering
        \includegraphics[width=10cm]{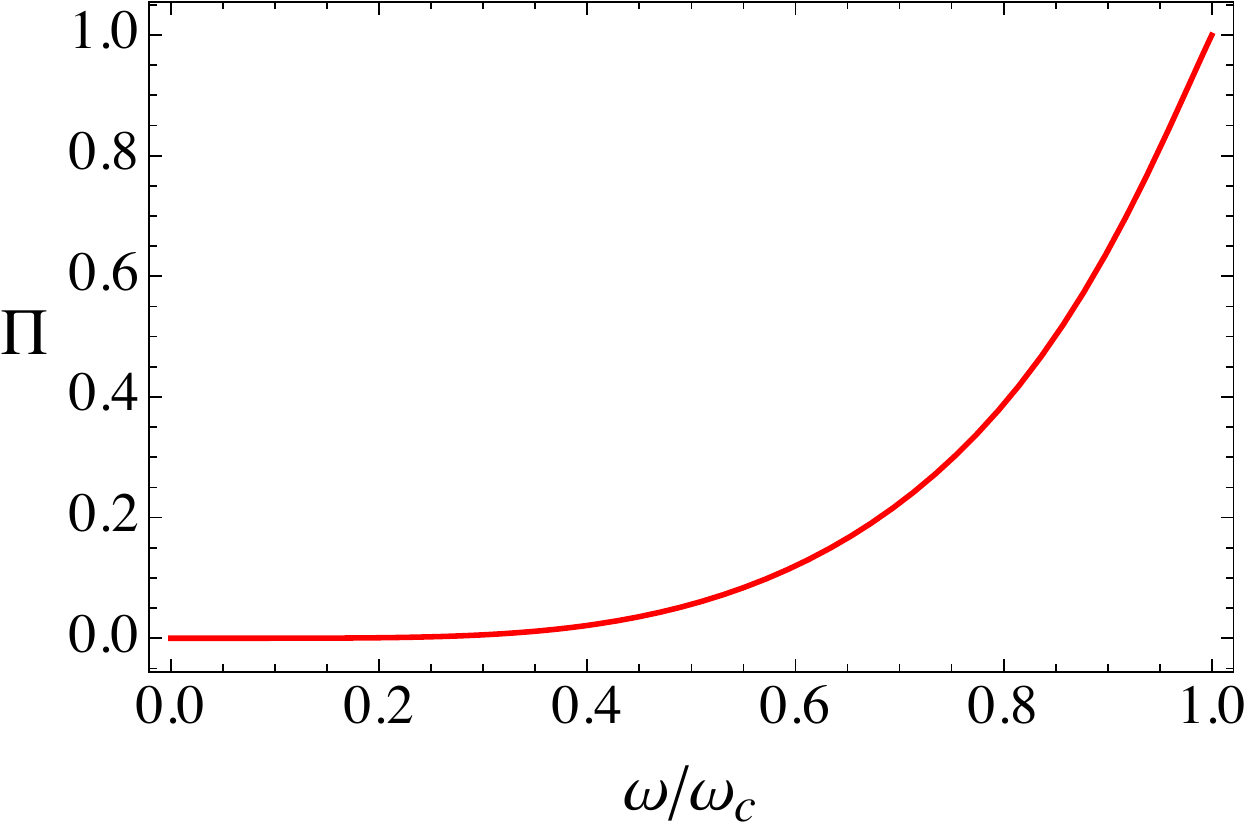}
        \vspace{-6pt}
        \caption{The plot of the degree of the circular polarization as a function of frequency normalized by the critical frequency. %Here, $\omega/\omega_{\rm c}$ is a dimensionless parameter, where $\omega_c$ is a critical value expressed as (4.2). 
        Here, $M_{\rm p}=10^{18}~\si{GeV}$, $\ell = 10^{8}~\si{km}$, $\eta = 1~\si{MeV}$, and $\lambda = 2$, thereby $\omega_{\rm c} = 10^{-21}~\si{eV}$ are used. The CS gravity is valid as long as $\omega/\omega_{\rm c} < 1$. 
        As $\omega/\omega_{\rm c}$ approaches the $1$, the degree of circular polarization increases. }
        \label{result}
        \vspace{0.7cm}
\end{figure}

We are now in a position to discuss implications for cosmology. First, we need to clarify the condition that a sizable circular polarization of GWs can be produced. From Eq.~(\ref{eq:eom_03}), we see that the CS correction to the Einstein gravity comes in the form of $\omega\ell^2\phi'/M_{\rm p}$. Since the CS gravity has to be small relative to the Einstein gravity, the CS correction has to satisfy
    \begin{eqnarray}
    \label{restriction}
    \frac{\omega\ell^2}{M_{\rm p}}\phi' = \frac{\ell^2}{M_{\rm p}}\omega\eta^2\sqrt{\frac{\lambda}{2}}\h  \mbox{sech}^2\left(\sqrt{\frac{\lambda}{2}}\eta z\right)< \frac{\ell^2}{M_{\rm p}}\omega \eta^2\sqrt{\frac{\lambda}{2}} < 1\,,
    \end{eqnarray}
where we used Eq.~(\ref{domain wall}). The critical frequency $\omega_c$ is obtained from the last inequality
\begin{eqnarray}
    \label{omegac}
   \omega_{c} = \frac{M_{\rm p}}{\ell^2 \eta^2}\sqrt{\frac{2}{\lambda}} \ .
\end{eqnarray}
In Figure \ref{result}, we plotted the degree of the circular polarization $\Pi$ for the parameters $\ell = 10^{8}~\si{km}$, $\eta = 1~\si{MeV}$, and $\lambda = 2$. We see that a sizable circular polarization can be produced. In particular, it becomes maximum when the frequency of GWs satisfies $\omega=\omega_c$, that is, when the CS correction gets comparable to Einstein gravity.  

The domain wall surface energy density is constrained as $\sigma \lesssim (\si{MeV})^3$ by the CMB power spectrum~\cite{Lazanu:2015fua}. Then we find  $\eta\sim{\rm MeV}$ at most by using Eq.~(\ref{sigma}). In~\cite{Babichev:2021uvl}, the authors consider a model where the domain wall tension decreases with the expansion of the universe and satisfies this constraint. If we assume the coupling constant $\sqrt{\lambda}\sim{\cal O}(1)$ and use $M_{\rm p}=10^{18}~{\rm GeV}$, the constraint of Eq.~(\ref{restriction}) is written as
    \begin{eqnarray}
    \label{constraint of ell}
    \omega\ell^2 < 10^{15}~\si{eV}^{-1}\ .
    \end{eqnarray}
Currently, measurements of frame-dragging effects around the earth by Gravity Probe B and the LAGEOS satellite constrain the characteristic CS length scale~\cite{Ali-Haimoud:2011zme} as
\begin{eqnarray}
\ell \lesssim 1 {\rm AU}\sim 10^8~{\rm km}\sim 10^{18}~{\rm eV}^{-1}\,.
\label{ell}
\end{eqnarray}
Suppose that $\ell\sim 10^{8}~{\rm km}$ then Eq.~(\ref{omegac}) gives $\omega_c\sim 10^{-21}~{\rm eV}\sim 1~\mu{\rm Hz}$. %Let us consider searching signals of the circular polarization of GWs around the frequency  $1~\mu{\rm Hz}$. 
We should observe circular polarization around $\omega_c\sim 1~\mu{\rm Hz}$. If the circular polarization was not observed, it would mean that $\ell$ is smaller than $10^8$ km. Repeating this procedure by decreasing $\ell$, we can get more stringent constraints on $\ell$. Note that decreasing $\ell$ corresponds to increasing $\omega_c$. If the circular polarization of GWs was observed at a frequency, say $\omega_c=1$ GHz,
it would imply the existence of axion and the CS length scale around $10$ km. Then, it will be important to probe such high frequency GWs~\cite{Aggarwal:2020olq,Ito:2019wcb,Ito:2020wxi,Ito:2022rxn,Ejlli:2019bqj,Domcke:2022rgu,Berlin:2021txa}.

Although we supposed that GWs are scattered by the domain walls at present, the scattering could have occurred in the early universe. If so, the GWs having passed through the domain wall should be red shifted at present. Then we need to take into account that the frequency $\omega$ of the GWs becomes higher in the past. 
Thus, we would be able to obtain a stronger constraint on $\ell$ using Eq.~(\ref{constraint of ell}). It would be also worth studying circular polarization of primordial GWs from the point of view of searching axion. Future space observations~\cite{Amaro-Seoane:2012aqc} would be desired for this purpose.

%==========================================================
%==========================================================

\section{Conclusion}
In this paper, we studied the circular polarization of GWs 
in an axion domain wall background. We first derived the EOM for the  GWs in CS gravity. We solved the EOM numerically and evaluate the degree of the circular polarization. We found that the degree of the circular polarization depends on the frequency of the GW and increases as the characteristic CS length scale becomes larger. When the effect of the CS correction becomes significant compared to Einstein gravity, the degree of the circular polarization becomes maximum. Depending on the value of the coupling strength of CS gravity, we may be able to observe the circular polarization that arose out of the axion domain walls. If it happened, observation of the circular polarization would provide us the information about CS gravity and a new avenue to search for axion.  

If the circular polarization of GWs is found in a large solid angle of the sky, the origin of the circular polarization could be a condensation of the axion that consists of the domain wall. 
To be consistent with CMB observations, we need to assume that  
the energy scale of the domain wall is less than 1~MeV. It turned out that observations of GWs with a frequency higher than $10^{-6}$ Hz is useful for getting information about CS gravity and axion. We argued that the characteristic CS length scale will be more constrained by detecting the circular polarization of the high frequency GWs. We also discussed that it would imply the existence of axion and the CS length scale around $10$ km if the circular polarization of GWs should be observed around $1$ GHz.

%==========================================================
\section*{Acknowledgments}
S.\ K. was supported by the Japan Society for the Promotion of Science (JSPS) KAKENHI Grant Number JP22K03621.
J.\ S. was in part supported by JSPS KAKENHI Grant Numbers JP17H02894, JP17K18778, JP20H01902, JP22H01220.
%==========================================================

\appendix
\setcounter{section}{0}
%\def\thesection{Appendix\h\Alph{section}}

%==========================================================
\section{Quadratic action of GWs in CS gravity}
%==========================================================
We derive the quadratic action of CS gravity (\ref{eq:q_action}). Up to the second-order in $h$, the metric an its inverse can be expressed as 
    \begin{eqnarray}
    \label{}
    g_{ij} &=& \delta_{ij}+h_{ij}  \ ,\\[4pt]
    g^{ij} &=& \delta^{ij}-h^{ij}+h^{il}h_{il}  \ .
    \end{eqnarray}
The second order perturbations of Chern-Simons term is given by
    \begin{eqnarray}
    \label{}
    \left[\sqrt{-g}\phi R\tilde{R}\right]^{(2)}
    &=&
    \frac{1}{2}\sqrt{-g}\phi\varepsilon^{\rho\sigma\alpha\beta}\left[R^{\mu\nu}{}_{\alpha\beta}R_{\nu\mu\rho\sigma}\right]^{(2)}\no\\[6pt]
    &=& 2\phi\vare^{ijk}\left(\left[2\RRa\right]^{(2)}+\left[\RRb\right]^{(2)}\right)
    \end{eqnarray}
where the superscript (2) denotes the second order perturbation. Since the background is Minkowski spacetime,  in the transverse traceless gauge, we obtain
    \begin{eqnarray}
    \label{a_01}
    \vare^{ijk}\left[2\RRa\right]^{(2)}
    &=&
    \vare^{ijk}\dot{h}^m{}_{k,j}\ddot{h}_{im} \ , \\[6pt]
    \label{a_02}
    \vare^{ijk}\left[\RRb\right]^{(2)}
    &=&
    \vare^{ijk}\dot{h}_{im,l}
    \left(h^{l}{}_k{}^{,m}{}_{,j}-h^{m}{}_k{}^{,l}{}_{,j} \right) \ .
    \end{eqnarray}
It is straightforward to show the following identity
    \begin{eqnarray}
    \label{a_05}
  &&  2\vare^{ijk}
    \left(\dot{h}^{m}{}_{k,j}\ddot{h}_{im}
    +\dot{h}_{im,l} h^{l}{}_k{}^{,m}{}_{,j}
    -\dot{h}_{im,l} h^{m}{}_k{}^{,l}{}_{,j}
        \right)\no\\[6pt]
    &&\qquad = 
    \vare^{ijk}\pa_t
        \left(
            \dot{h}^m{}_{k,j}\dot{h}_{im}
            +h_{im,l}h^{l}{}_{k}{}^{,m}{}_{,j} 
            -h_{im,l}h^{m}{}_{k}{}^{,l}{}_{,j}  
        \right)   \nonumber \\[6pt]
    &&\qquad\qquad    +
        \vare^{ijk}\pa_j
        \left(
            \ddot{h}^m{}_i\dot{h}_{km}
           +\dot{h}_{im,l}h^{l}{}_{k}{}^{,m}
            -\dot{h}_{im,l}h^{m}{}_{k}{}^{,l} 
        \right).
    \end{eqnarray}
The reason that the action is expressed in terms of total derivatives is that the CS term is a topological term. In other words, the CS term appears when there is a coupling field $\phi$.
Thus, the quadratic action reads
    \begin{eqnarray}
    \label{a_06}   
    \overset{(2)}{S_{CS}}&=&
    \frac{M_p \ell^2}{8}\int d^4x \left[\sqrt{-g} \phi R\tilde{R}\right]^{(2)}\no\\[6pt]
    &=&\frac{M_p \ell^2}{8}
    \int d^4x\left[
    \phi\vare^{ijk}\pa_t
        \left(
            \dot{h}^m{}_{k,j}\dot{h}_{im}
            +h_{im,l}h^{l}{}_{k}{}^{,m}{}_{,j} 
            -h_{im,l}h^{m}{}_{k}{}^{,l}{}_{,j}   
        \right) \right.\nonumber\\[6pt]
    && \left. \qquad \qquad \qquad +
        \phi\vare^{ijk}\pa_j
        \left(
           \ddot{h}^m{}_i\dot{h}_{km}
           +\dot{h}_{im,l}h^{l}{}_{k}{}^{,m}
            -\dot{h}_{im,l}h^{m}{}_{k}{}^{,l}
        \right)\right]\no\\[6pt]
    &=&-\frac{M_p \ell^2}{8}\int d^4x
        (\pa_z\phi)\vare^{izk}
        \left(
            \ddot{h}^m{}_i\dot{h}_{km}
           +\dot{h}_{im,l}h^{l}{}_{k}{}^{,m}
            -\dot{h}_{im,l}h^{m}{}_{k}{}^{,l}
        \right),
    \end{eqnarray}
where we used the fact $\phi$ depends  only on $z$. 
%
%
%==========================================================
\section{Equation for GWs in CS gravity}
%==========================================================
Now, we derive EOM from the quadratic action. 
Taking the variation  of the action (\ref{eq:q_action})
with respect the metricpertubation $h_{ij}$, we 
obtain
    \begin{eqnarray}
    \label{}   
    \Box h^{ij}
    =
    -\frac{ \ell^2}{2M_p}\varepsilon^{izk}
    \left( \pa_z \phi \ \Box \dot{h}^j{}_{k}
    +   \pa^2_z \phi \ \pa_z\dot{h}^j{}_{k}
    \right)  + (i\leftrightarrow j)
    \end{eqnarray}
 or    
    \begin{eqnarray}
    \Box h^{ij}
    &=&
   \frac{ \ell^2}{2M_p} \varepsilon^{zik}
    \left(\phi'\Box
    +\phi''\pa_z
    \right)\dot{h}^j{}_{k}
     + (i\leftrightarrow j)\h ,
    \end{eqnarray}
where $\Box$ denotes the d'Alembert operator, and prime denotes the derivative with respect to $z$.
%
%
%
%==========================================================
\section{Effective potential}
%==========================================================
We  derive the effective potential (\ref{V_eff}).  Substituting $H(z)_{R/L}=x(z)f(z)_{R/L}$ into the Eq.~ (\ref{eq:eom_03}), we obtain,
    \begin{eqnarray}
    \label{c_01}
    f''+\left(2\frac{x'}{x}+\frac{F'}{F}\right)f'+\left(\frac{x''}{x}+\frac{F'}{F}\frac{x'}{x}+\omega^2\right)f = 0,
    \end{eqnarray}
where $F\equiv 1\pm\omega\ell^2\phi'/M_{\rm p}$ and the indices ${R/L}$ are omitted. In oder to transform (\ref{c_01}) into the form of Schr\"{o}dinger type equation, the coefficients of $f'$ must vanish, hence we have
    \begin{eqnarray}
    \label{c_02}
    \frac{x'}{x}=-\frac{1}{2}\frac{F'}{F}.
    \end{eqnarray}
Without loss of generality, this can be solved as $x=1/\sqrt{F}$.
Substituting (\ref{c_02})  into (\ref{c_01}), we obtain
    \begin{eqnarray}
    \label{c_04}
	\left[-\frac{\pa^2}{\pa z^2}-\frac{1}{4}\left(\frac{F'}{F}\right)^2+\frac{1}{2}\frac{F''}{F}\right]f 
    =\omega^2 f_.
    \end{eqnarray}
%
%==========================================================
\printbibliography
%==========================================================
\end{document}